\documentclass[conference]{IEEEtran}
\pdfoutput=1
\IEEEoverridecommandlockouts
\usepackage{booktabs,ragged2e}
\usepackage[flushleft]{threeparttable}

\usepackage{tabularx}
\usepackage{multirow}
\newcolumntype{L}{>{\raggedright\arraybackslash}X}
\usepackage{cite}
\usepackage{amsmath,amssymb,amsfonts}
\usepackage{algorithmic}
\usepackage{graphicx}
\usepackage{textcomp}
\usepackage{xcolor}
\def\BibTeX{{\rm B\kern-.05em{\sc i\kern-.025em b}\kern-.08em
    T\kern-.1667em\lower.7ex\hbox{E}\kern-.125emX}}
\begin{document}

\title{A Technical Review of Wireless security for the Internet of things:  Software Defined Radio perspective\\
}

\author{\IEEEauthorblockN{ Jose de Jesus Rugeles Uribe}
\IEEEauthorblockA{\textit{Applied Sciences PhD program} \\
\textit{Universidad Militar Nueva Granada}\\
Bogotá, Colombia \\
jose.rugeles@unimilitar.edu.co}
\and
\IEEEauthorblockN{Edward Paul Guillen}
\IEEEauthorblockA{\textit{Telecommunications Engineering program} \\
\textit{Universidad Militar Nueva Granada}\\
Bogota, Colombia \\
edward.guillen@unimilitar.edu.co}
\and
\IEEEauthorblockN{Leonardo S. Cardoso}
\IEEEauthorblockA{\textit{Univ Lyon, INSA Lyon} \\
\textit{Inria, CITI}\\
Lyon, France \\
leonardo.cardoso@@insa-lyon.fr}
\and

}

\maketitle

\begin{abstract}
The increase of cyberattacks using IoT devices has exposed the vulnerabilities in the infrastructures that make up the IoT and have shown how small devices can affect networks and services functioning. This paper presents a review of the vulnerabilities of the wireless technologies that bear the IoT and assessing the experiences in implementing wireless attacks targeting the Internet of Things using Software-Defined Radio (SDR) technologies. A systematic literature review was conducted. The types of vulnerabilities and attacks that can affect the wireless technologies that stand the IoT ecosystem and SDR radio platforms were compared.  On the IoT system model layer, perception layer was identified as the most vulnerable. Most attacks at this level occur due to limitations in hardware, physical exposure of devices, and heterogeneity of technologies. Future cybersecurity systems based on SDR radios have notable advantages due to their flexibility to adapt to new communication technologies and their potential for the development of advanced tools. However, cybersecurity challenges for the Internet of Things are so complex that it is needed to merge SDR hardware with cognitive techniques and intelligent techniques such as deep learning to adapt to rapid technological changes.

\end{abstract}

\begin{IEEEkeywords}
Internet of Things, Software Defined Radio(SDR), Cybersecurity, Radio communication, Cyberattack, Vulnerabilities
\end{IEEEkeywords}

\section{Introduction}
On September 5 and 6, 2019 a cyber-attack affected several Wikipedia sites in Europe - including  Germany, France,Italy and Poland, at the same time as parts of the Middle East \cite{Bankinfosecurity2019}.IoT devices were used to take down Wikipedia through distributed denial of service (DDoS) attacks. The IoT were used to amplifying the potential cyberattack surface.According to the World Economic Forum report, attacks on IoT devices grew by more than 300\% in the first half of 2019 \cite{Weforum2020},\cite{Kaspersky2019}.

On May , 2018 the VPNFilter malware targets at least 500K networking devices  in 54 countries \cite{Enisa2018}. Home-office network devices, as well as network-attached storage devices were infected. VPNFilter exploited various vulnerabilities in several models and brands of routers and network-attached storage (NAS) devices.
The malware capabilities identified including data exfiltration, spying on traffic, and rendering the infected device unbootable. The malware code overlaps with versions of the BlackEnergy malware, which was responsible for multiple large-scale attacks that targeted devices in Ukraine.








On April 7, 2017, a cyberattack on the emergency system in the city of Dallas, in the United States, triggered 156 emergency sirens at 11:40 p.m., causing panic and fear of a possible terrorist attack \cite{WIRED2017}.The case study carried out by the company Bastille Networks \cite{BalintSeeberBastille2018}, found that the attacker exploited the vulnerability of the wireless infrastructure making up the emergency system to control its operation and also found that the United States has over 5000 similar emergency systems distributed throughout different cities, universities, military facilities and industries.

On October 21, 2016 \cite{DYN2016}, a denial-of-service cyber-attack targeting Internet provider DynDNS caused problems for sites such as Twitter, Netflix, Spotify, Airbnb, Reddit, Etsy, SoundCloud and the New York Times, affecting much of the United States of America. According to subsequent analyses \cite{Securityintelligence},thousands of IoT devices such as cameras and home routers were used, which were violated in various ways. The traffic volume of this attack has been the highest recorded so far: 1.2 Tbps; a similar event recorded previously had reached 620 Gbps.

These security incidents are an exemplify the vulnerability of some of today's communication systems. At a point in history where society is moving rapidly towards technological scenarios supported largely by the Internet of Things (IoT) communications infrastructure \cite{Gupta2015}, such threats represent a global challenge. 

The International Telecommunication Union estimates 25 billion connected devices for 2020 \cite{ITU2016}; and according to the Ericsson mobility report \cite{Ericcson2017} by 2022 the number of connected devices will be 29 billion, with a projected increase of 21\% between 2016 and 2022.\cite{Stellios2018} identifies potentially vulnerable scenarios and describes risks in critical infrastructure such as power distribution, SCDA, smart transport, medical care and smart home automation. When analyzing the technological infrastructure that supports the IoT \cite{Zayas2017}, we find a diverse ecosystem of wireless technologies, with rapid growth and some gaps in the regulation of device-manufacturing processes.

Risk assessment and analysis of the impact of cyberattacks is a concern that goes beyond technical and academic scenarios. The OECD (Organisation for Economic Cooperation and Development), in its Recommendation on Digital Security Risk Management for Economical and Social Prosperity \cite{Economy2015}, analyses the issue of digital security and its impact on the economic development of countries.\cite{E&Y2015} estimates that 70\% of connected IoT devices have vulnerabilities.  \cite{CommitteeonScienceSpace} analyses the risks and the ways in which such events can affect the operation of vital systems for people. In 2014, it was estimated that the economic impact of cybercrime was 500 billion dollars. By 2018, this figure had reached 600 billion \cite{Lewis2018}.In 2021,cybercrime damages might reach US\$6 trillion \cite{Weforum2020}.

Some Governments, concerned about threats to their communications infrastructures, have created special programs to counter them. The United States government through the Advanced Research Projects Agency (DARPA) created the Trusted Integrated Circuits program to develop technologies that ensure reliability in the manufacturing processes of electronic devices used in military systems. Other countries are undertaking similar initiatives through cyber security agencies such as Agence nationale de la sécurité des systèmes d'information (ANSSI) in France; Nationales Cyber-Abwehrzentrum in Germany; and Defence Cyber Operations Group in the United Kingdom.

A large part of IoT technologies are wireless and form an ecosystem of radios designed from various communication protocols \cite{Frustaci2018}. The old radio systems required specific hardware to operate in a very limited range of frequencies \cite{Mitola1999}. Modern radios are evolving towards Software-defined radio technology (SDR), where the same hardware platform can adapt and become a transmitter and/or receiver system operating under various technologies, by modifying the software's configuration parameters in the device. Although the development of ADC-DAC converter manufacturing techniques and FPGA-based processing systems have led to significant improvements in hardware performance, there are still several technical limitations to the production of the ideal SDR radio device proposed by Joseph Mitola in 1999 \cite{Mitola1999}. \cite{Pawelczak2011} presents an analysis of the first ten years of SDR technology development, while \cite{Mitola2015}  describes the evolution of trends in SDR considering the last two decades.

The large growth of wireless devices for IoT development also causes risks and vulnerabilities to increase. The types of wireless attacks are varied and there is no single classification. Among the best known are: spoofing, eavesdropping, sidechannel, jamming, replay and spoofing. All of them are based on exploiting the various vulnerabilities present in radio devices or in their communication protocols.

The review made allows, firstly, to clarify the dimension of the wireless ecosystem of the IoT, with its technologies and protocols, and then the concepts related to radio technology defined by software and the types of existing hardware and software resources, as well as their characteristics and technical possibilities. It then addresses the concepts of wireless cybersecurity, in order to finally identify the types of vulnerabilities and attacks that can affect the wireless technologies that make up the IoT ecosystem SDR radio platforms and cognitive radio networks. In the document, the security aspects related to SDR technology are analyzed in two ways: first by identifying the vulnerabilities inherent to SDR devices since the flexibility and reconfigurability of SDR hardware make its adaptability very high, but they can also be affected by attacks capable of modifying radio behavior. This aspect is very important for the future of wireless systems, where cognitive radio networks will be supported by reconfigurable radios. The second aspect considered in the review was the classification of various experiences in implementing wireless attacks using commercial SDR radios, and the use of these platforms as tools for the analysis of risks and vulnerabilities in wireless systems. The contributions of the review are the following:

\begin{itemize}[]
    \item Classifying and identifying the technologies that make up the wireless ecosystem of the Internet of Things, their standards and operating frequencies.

    \item Contextualizing definitions and concepts related to SDR technology and the various hardware architectures.

    \item Classifying the most common SDR hardware platforms considering their main technical characteristics.

    \item Identifying the types of SDR radios and cognitive radio networks vulnerabilities.
    
    \item Identifying the most common types of wireless vulnerabilities and IoT wireless ecosystem vulnerabilities.
    
    \item Understanding wireless attack implementation experiences and vulnerability assessments for wireless technologies, and identifying the types of SDR radio platforms used.
\end{itemize}

The document is organized as follows: section two presents the context of the technologies that make up the wireless ecosystem of the Internet of Things, their operating frequencies and standards. Thereafter, section three includes definitions of concepts related to software-defined radio, the types of hardware architectures and a technical comparison of the most commonly used SDR platforms. 
The fourth section addresses issues related to wireless cybersecurity. This includes classifying SDR radio vulnerabilities, common wireless vulnerabilities, and IoT wireless ecosystem vulnerabilities and cognitive radio network vulnerabilities. This section also describes various experiences in wireless attacks and wireless security deployments, and the analysis performed using software-defined radio platforms. Experiences are identified and classified according to the types of attacks implemented and SDR platforms used are described. Finally, section five includes an analysis on the most significant elements identified from the review process carried out.

\section{WIRELESS ECOSYSTEM}
\begin{figure*}[ht!]
        \centering
        \caption{Ecosystem of wireless technologies used for the IoT. Based on \cite{KeysightTechnologies2019}, Fig 2}
        \includegraphics[width=1\textwidth]{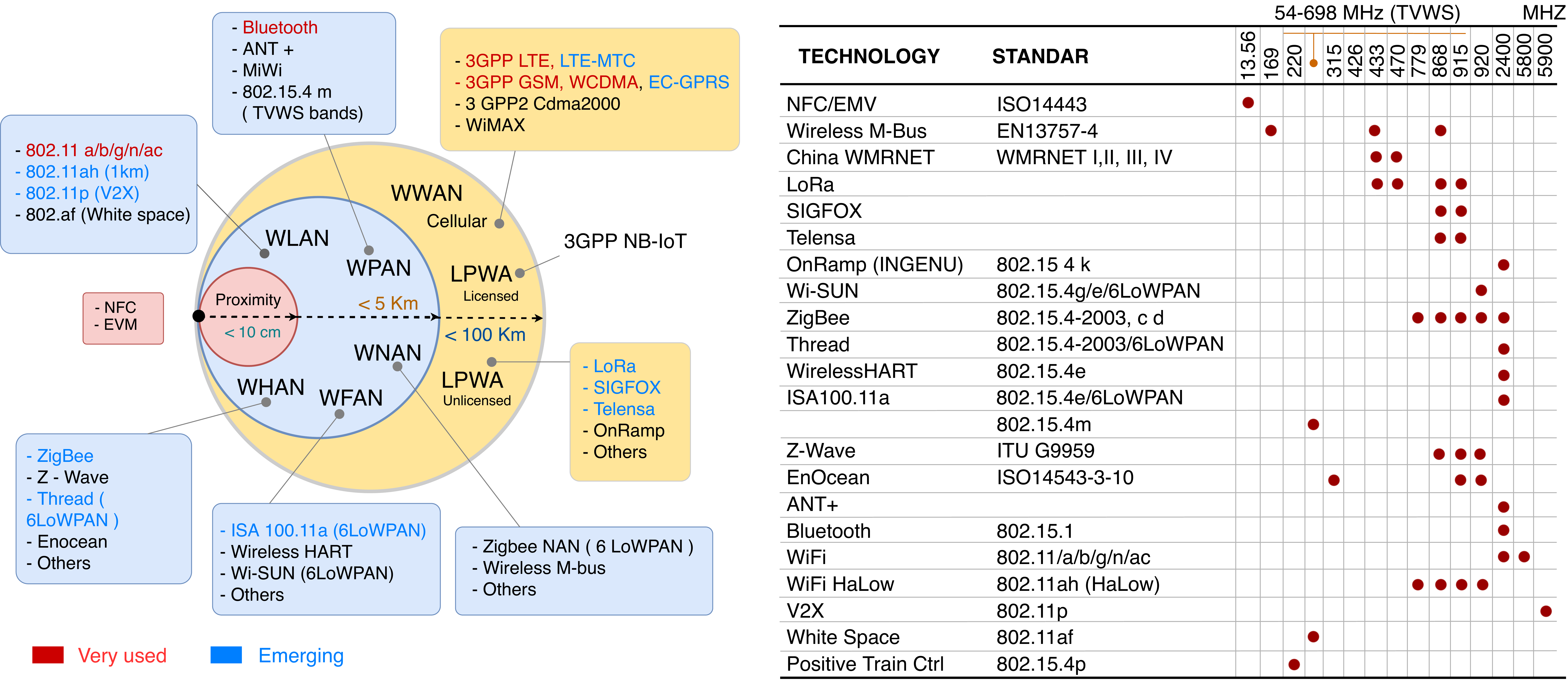}
        \label{fig:ecosistema}
\end{figure*}

Figure \ref{fig:ecosistema} shows a classification based on \cite{KeysightTechnologies2019} for some of the wireless communication technologies used in the implementation of the Internet of Things, taking into account the range of their scope. By classifying wireless technologies by range, we can talk about technologies: of proximity, WPAN, WFAN, WLAN, WNA and WWAN.

The frequency bands for the operation of each can be appreciated, along with the corresponding standards. This technological scenario considers the technologies existing in 2019 and their evolution towards a 5G technologies-based future ecosystem. \cite{Yang2018},\cite{Palattella2016},\cite{Spiridon} discuss several aspects related to the evolution of these scenarios, their challenges, possibilities and limitations. This set of technologies or ecosystem has enabled the development of concepts such as Internet of Things, Smart Cities or Smart grids. Some of them, such as Bluetooth, 3GPP GSM,802.11g, NFC, have been massified by their use in devices such as mobile phones, computers, access control systems, sound systems, payment cards, etc., while others such as ZigBee, Lora,802.11p, are regarded as emerging technologies. In this context, with the rapid development of the wireless ecosystem, enormous challenges arise, such as opportunistic use of spectrum, regulatory matters, interoperability between technologies, quality of service and a key element: system security \cite{Frustaci2018}.

\section{SOFTWARE DEFINED RADIO}
In 2009, the ITU-R International Telecommunication Union Working Group 1B defined the term SDR (Software-Defined Radio) for the 2012 World Radio communication Conference \cite{ITU-R2009}, as: “A radio transmitter and/or receiver employing a technology that allows the RF operating parameters including, but not limited to, frequency range, modulation type, or output power to be set or altered by software, excluding changes to operating parameters that occur during the normal pre-installed and predetermined operation of a radio according to a system specification or standard.” Already in 1992, Joseph Mitola III \cite{MitolaJ1999} had defined the term software Radio as: “A software Radio is a radio wave channel modulations waveforms are defined in software.That is, waveforms are generated as sampled digital signals, converted from digital to analog via wideband DAC (Digital Analog Converter) and the possibly unconverted form IF (Intermediate Frequency) to RF (Radio Frequency) .The receiver, similar, employs a wideband ADC (Analog to Digital Converter) that captures all the channels of the radio node software. The receiver then extracts, down converts and demodulated the channel waveform using software on a general purpose processor.”

\subsection{SDR Hardware Platforms}

The concept of SDR platforms arose with Mitola in 1992 \cite{Mitola1993}.In 1996, at MIT, the \textit{SpectrumWare software radio project} was carried out, where a prototype receiver of a GSM base station was developed \cite{Tennenhouse1996}.In 1999, SDR SpeakEasy I/II radios were manufactured, for military use, and financed by the US government.In 2004,  the company Ettus was created and the production of SDR-USRP hardware began; its use has expanded and has become one of the most widely used platforms nowadays \cite{Machado2015},\cite{Pawelczak2011},\cite{Mitola2015}. 

Table \ref{tab:ADC-DAC} presents a summary of the technical characteristics of some commercially available software-defined radio platforms: The number of bits used by analog digital and/or digital analog converters is regarded (DAC), as well as their sampling rates, bandwidth and transmission, and/or reception possibilities.

The increase in the sampling capabilities of converters attests to the evolution of SDR platforms, which enables the implementation of modern wireless communications standards \cite{Machado2015}. SpeakEasy used converter devices with sampling rates in the order of 200 Kb/s for its manufacture in 1999 \cite{Cook1999}.The USRP1, designed in 2004, has an ADC converter sampling rate of 64 [MS/S], compared with the reference USRP X310 released to the market in 2016; the sampling rate is observed to have increased to 200 [MS/s]. The evolution of the platforms has also allowed the increase of radios' frequency range. As can be seen in the  table \ref{tab:ADC-DAC} the frequency ranges reach 6 GHz.
  \begin{table}[h!]
   \begin{tabular} { m{10em} p{3em} m{3em} m{3em} m{4em} }
    
        \hline
      \textbf{Platform} & \textbf{ADC /DAC [Bits]} & \textbf{ADC /DAC [MS/s]} & \textbf{Tx/Rx} & \textbf{Fmin-Fmax [MHz]}\\
        \hline
        RTL \cite{RTL-SDR.COM} & 8/- & 3.2/- & Rx & 25-1750  \\
        \hline
        FUNCube \cite{FUNcube} &  16/- & 96 KHz  & Rx & 64-1700  \\
        \hline
        Airspy-mini \cite{Airspy} & 12/- & 20/- & Rx & 24-1700  \\
        \hline
        HackRF One \cite{HackRFOne} &  8/10 & 20/20 & Tx-Rx & 1-6000  \\
        \hline
        Pluto \cite{AnalogDevices} &  12/12 & 61.4 /61.4 & Tx-Rx & 325-3800 \\
        \hline
        BladeRF 2.0 Micro \cite{Nuand} & 12/12 & 61.4 /61.4 & Tx-Rx & 47-6000  \\
        \hline
        USRP-1 \cite{EttusResearchc}& 12/14 & 64/128 & Tx-Rx & DC-6000  \\
        \hline
        Nutaq PicoSDR \cite{Nutaq}&  12/12 & 80/80 & Tx-Rx & 56-6000  \\
        \hline
        USRP-2 \cite{EttusResearchc}& 14/16 & 100/400 & Tx-Rx  & DC-6000  \\
        \hline
        USRP-N210 \cite{EttusResearchc}& 14/16 & 100/400 & Tx-Rx  & DC-6000  \\
        \hline
         WARP-V3 \cite{MangoCommunications} &  12/12 & 100/170 & Tx-Rx & 2400/5000  \\
        \hline
        LimeRF LMS7002M \cite{LimeMicrosystems} &  12/12 & 160/640 & Tx-Rx & 0.1-3800  \\
        \hline
        USRP X310 \cite{EttusResearchc}&  14/16 & 200/800 & Tx-Rx & DC-6000  \\
        \hline
        USRP N310 \cite{EttusResearchc} &  16/14 & 153.6 /153.6 & Tx-Rx & 10-6000  \\
        \hline
        AIR-T \cite{Digital} &  12/10 & 245.7 /245.7 & Tx-Rx & 300-6000  \\
        \hline
        CRIMSON \cite{Pervices} &  16/16 & 370,16 /2500,16 & Tx-Rx & 0.1-6000  \\
        \hline     
      \end{tabular}
      \\
      \caption{\label{tab:ADC-DAC} Software Defined radio platforms comparison}

    \end{table}

In addition to improvements in the features of radio interfaces, SDR hardware has also evolved in increasing its computational capacity by integrating DSP and FPGAs devices into its systems.The \textit{Xilinx Virtex 6} is used in SDR devices such as PicoSDR, WARP v3; the X-Series USRP uses \textit{Xilinx Kintex 7} FPGAs and platforms such as CRIMSON \cite{Pervices} integrate an \textit{FPGA-Arria V ST SOC} and an \text{ARM Cortex-A9 MP}. The AIR-T(\textit{Artificial Intelligence Radio - Transceiver }) platform \cite{Digital} combines a 2x2 SDR MIMO system with a 256-core  \textit{GPU Jetson TX2} on a single card.Its ADC and DAC converters reach sample rates of up to 245.7 [MS/s]. Other platforms can be classified in the low-cost hardware category such as RTL \cite{RTL-SDR.COM}, HackRF  \cite{HackRFOne},AirSpy \cite{Airspy},FUNCube \cite{FUNcube},BladeRF \cite{Nuand}, LimeRF \cite{LimeMicrosystems} and Pluto \cite{AnalogDevices}; These devices, along with free software tools such as GNU Radio and driver integration for SDR devices signal processing platforms such as Matlab and Labview have enabled the development of software-defined radio technology worldwide.

\section{CYBERSECURITY FOR WIRELESS TECHNOLOGIES }
To assess the vulnerability level of a system or technology, the \textit {National Infrastructure Advisory Council (NIAC)} \cite{NIAC} established the \textit{Common Vulnerability Score System} (CVSS) \cite{CVSS}.  \cite{Qu2016} uses this classification to perform a vulnerability assessment on Bluetooth low energy technology in IoT systems. Moreover, the U.S. Department of Homeland Security, through the Office of Cybersecurity and Communications (CSC), set up a mechanism to identify and classify attack patterns through a public catalog known as CAPEC (\textit{Common Attack Pattern Enumeration and Classification})\cite{CAPEC}. 

\subsection{MOST COMMON WIRELESS VULNERABILITIES}

According to \cite{Fragkiadakis2013}, the security basics for the operation of a wireless network are confidentiality, integrity, availability and access control. Confidentiality must ensure that network data cannot be read by unauthorized users, while integrity detects intentional or accidental changes to data being carried over the network. Availability ensures that devices and users can use the network and its resources when they need it, and access control restricts resources to authorized users only. 

The company Bastille Networks conducted a classification of the top 10 vulnerabilities of wireless systems \cite{BastilleNetworks}.Table \ref{tab:bastille} describes each one. 


\begin{table}[h!]
\small
    \begin{tabular}{p{2cm}p{6cm}}
    \hline
    \textbf{Vulnerability} & \textbf{Description}  \\
    \hline 
    \textit{Rogue cell towers} & It use IMSI catchers or Stingrays. A mobile cell allow spoof an cellular communication. The attacker can heard and read SMS.\\
    \hline
    \textit{Rogue WiFi Hotspots} & The WiFi hotspots can be used to deploy man in the middle attacks. Is possible keep watch on network traffic or stole user's credentials.\\
    \hline
    \textit{Bluetooth Data exfiltration} & It uses a mobile device with Bluetooth that avoids network controls by means of internet access across the cellular network.\\
    \hline
    \textit{Eavesdroping / surveillance devices} & Eavesdropping devices voice-activated, with FM or GSM transmission. Devices are hidden in offices or meeting rooms.\\
    \hline
    \textit{Vulnerable wireless peripherals} & keystroke injection attacks. It uses the weakness of wireless keyboards and/or mouses without data encryption.\\
    
    \hline
    \textit{Unapproved cellular device presence} & Unapproved cellular device use in restricted areas. \\
    \hline
    \textit{Unaproved wireless cameras} & Unapproved wireless cameras are a security breach.\\
    \hline
    \textit{Vulnerable wireless building controls} & Home automation devices with  by default unsecured configurations.\\
    \hline
    \textit{Unapproved IoT Emitters}& Devices like thermostats or wireless sensors inside the buildings using Technologies like Zigbee or LoRa with long-range coverage.\\
    \hline
    \textit{Vulnerable building alarm systems} & 
    Security systems elements like door sensors, motion detectors can be sensitive to jammer attacks using software defined radios \\.
    
    \end{tabular}
    \caption{Wireless vulnerabilities according to the Bastille Networks classification \cite{BastilleNetworksb}}
    \label{tab:bastille}
\end{table}

\subsection{SDR radios vulnerabilities}

The types of wireless attacks are varied, \cite{Moura2012} presents a classification of attacks targeting SDR tactical radios, while \cite{Fragkiadakis2013}, considers attacks with SDR radios within the Cognitive Radio Network Vulnerability Classification (CRNs). \cite{Soliman2018} analyzes the taxonomy of threats to cognitive radio networks by taking into account the layers of the OSI model for classification. Other studies focus specifically on the detailed analysis of a specific type of attack and/or vulnerability, such as \cite{Lichtman2016} where jamming-type attacks are analyzed.

The applicability of the reconfigurable radios concept has been proven in areas as demanding as tactical communications \cite{Moura2012}.However, there are technical factors such as power consumption, which restrict for their implementation to mobile devices. However, one of the vital aspects for their massification is to ensure that systems based on SDR technologies are protected against malicious codes \cite{Ulversoy2010}. 

The role of software-defined radio technology can be analyzed in several ways: firstly by considering the vulnerabilities of SDR radio platforms that can be exploited to affect their operation. A second approach arises from the use of SDR hardware for the generation of wireless attacks, which can affect the operation of various types of radio communication infrastructure, taking advantage of the vulnerabilities in protocols or technologies of the various radio ecosystems. The third aspect has to do with the potential of SDR technology for the development of diagnostic and control tools for cyber-threats and wireless cyber-attacks. 

\begin{figure*}[t]
        \centering
        \caption{Wireless attack classification, based on \cite{Moura2012}}
        \includegraphics[width=0.95\textwidth]{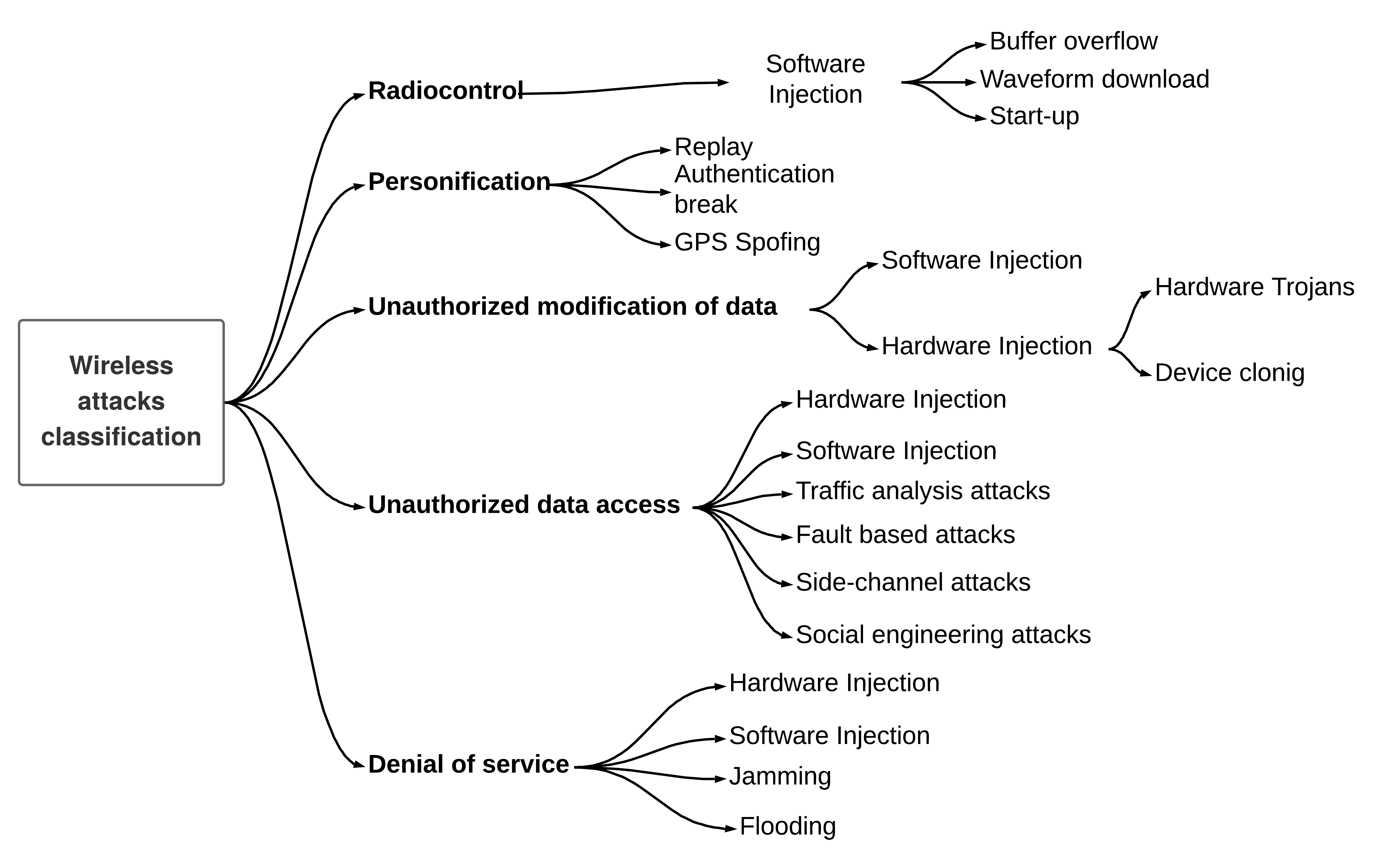}
        \label{fig:ataques_c1}
\end{figure*}

Figure \ref{fig:ataques_c1} presents a classification of the types of wireless attacks targeting software-defined radios taking into account the work done by \cite{Moura2012}. The study considers vulnerabilities in tactical radios with software-defined radio technology. The authors present five different types of attacks: radiocontrol, personification, unauthorized data modifications, unauthorized data access and denial of service. Each of these typologies group a set of types of attacks and each of them arises from exploiting some vulnerabilities present in the system. Each of the elements of this structure is described in detail below.

\subsection{IoT wireless ecosystem  vulnerbilities }
Table \ref{tab:ataques_iot} shows the classification of vulnerabilities of the Internet of Things according to a layer model described in \cite{Frustaci2018}.The three-level model considers: perception, transport and application. In the perception layer are physical IoT devices; various types of transducers obtain the data and devices are collected through wireless ecosystem technologies, such as those shown in Figure \ref{fig:ecosistema}. Generally, the devices used do not have high computing power due to limitations in power consumption or manufacturing costs.The transport layer is responsible for transmitting the information obtained in the perception layer to some of the processing systems, using access networks such as 3G, WiFi, ad hoc or the Internet. The application layer provides the services required by users. At this level, solutions such as smart-cities, smart-healthcare, etc are implemented. An application sublayer allows to manage all types of derivative services such as performing intelligent computing, implementing localization services, or serving as an interface with cloud computing systems. 

At each level, the authors identify a set of vulnerabilities. According to the analysis, it is considered that the level of perception is the least secure due to the physical exposure of IoT devices, hardware limitations and heterogeneity of technologies and where the following types of attacks can occur: physical, impersonation, denial of service ,and routing and data transit.
 
\begin{table}[ht!]
\small
\centering
\begin{tabular}{@{}ll@{}}
\toprule
\textbf{Layer} & \textbf{Vulnerability} \\ 
\midrule
\multirow{3}{*}{\textit{Application}} 
 & {Data jailbreak} \\
 & {Denial-of-service attacks} \\
 & {Malicious code injection } \\
 \hline
\multirow{3}{*}{\textit{Transport}} 
 & {Routing attacks} \\
 & {Denial-of-service attacks} \\
 & {Data transit attacks} \\
  \hline
\multirow{5}{*}{\textit{Perception}} 
 & {Physical attacks} \\
 & {Spoofing} \\
 & {Denial-of-service attacks} \\
 & {Routing attacks} \\
 & {Data transit attacks}\\
  \hline
\end{tabular}
\caption{\label{tab:ataques_iot}IoT Wireless Vulnerabilities, according to \cite{Frustaci2018}}
\end{table}

Perception layer attacks may be: physical, impersonation, denial of service, routing, and data traffic attacks.





\subsection{Attacks on Cognitive radio networks}

Table  \ref{tab:ataques_cognitive} shows a classification for the types of attacks on cognitive radio networks according to the classification made by \cite{Fragkiadakis2013}. The vulnerabilities of cognitive radio systems relate to two of the main characteristics in this technology: its cognitive capacity and its reconfigurability. In the first case, attacks seek to emulate primary transmitters and/or transmit false observations of the spectrum sensing process. In the second scenario, by installing a malicious code, an attacker can take control of the radio. 

\begin{table}[ht!]
      \centering
       \begin{tabular} { m{8em} m{4em} m{12em} } 
        \hline
       \textbf{Type of attack} & \textbf{Layer} & \textbf{Description }\\
        \hline
     Primary user emulation & Physical & Signals emulation of primary transmitter \\
        \hline
     spectrum sensing data falsification ,(SSDF) \cite{Sharifi2016}  & Physical & Wrong observations concerning to spectrum sensing \\
        \hline   
     Common control channel & Medium access  & Spoofing ,  congestion,jamming    \\
        \hline
     Beacon falsitication & Medium access & Disruption of synchronization between IEEE 802.22 WRANs\\ 
          \hline
   Cross-Layer & All layers & Advanced attacks across multiple layers \\ 
          \hline
     Software Defined Radio  & All layers & Manipulation of hardware and/or software on Software Defined Radios \\
        \hline
      \end{tabular}
      \caption{\label{tab:ataques_cognitive}Attacks on cognitive radio networks, based on \cite{Fragkiadakis2013}}
     
    \end{table}

Cognitive networks, by their nature, are also exposed to the vulnerabilities and attacks inherent to wireless networks and cognitive radios, for they are built using SDR radios and that makes them vulnerable to the same type of threats to which software-defined radio devices are exposed. 

\subsection{Cybersecurity experiences with SDR }

Table \ref{tab:ejemplos_SDR} presents a compendium of cases where SDR technology has been used to perform cybersecurity analyses on wireless technologies such as GSM, LTE, DECT, RFID, ACARS, ADS-B, LoRa, BLE, IEEE 802.11, IEEE 802.15.4, NFC or on communication systems with drones and vehicles; some of them are part of the wireless ecosystem of the Internet of things.The table specifies the type of SDR radio platform used and the type of attack or vulnerability implemented.

\begin{table*}[ht!]
\small
      \centering
       \begin{tabular} { m{17em} m{15em} m{10em} m{4em} } 
       
        \hline
       \textbf{Vulnerability, attack or analysis} & \textbf{Technology, objective} & \textbf{Hardware } & \textbf{References}\\
        \hline
       Spoofing & Drones & USRP & \cite{Nguyen}  \\
        \hline
       Spoofing & ACARS, FANS1/A & USRP B200 & \cite{Bresteau2018} \\
          \hline
       Spoofing & Vehicular Security (TMPS protocol) & USRP N210/ & \cite{Kilcoyne2016} \\
          \hline
       Spoofing & FM-based indoor localization. & USRP B100/WBX & \cite{Li2016} \\
          \hline
       GPS Spoofing & GNSS & USRP N210  & \cite{Schmidt2018} \\
          \hline
       Man in the middle & IoT(Bluetooth 2.1)  & USRP 2 & \cite{Barnickel2012} \\
           \hline       
       Man in the middle & GSM & USRP B200 & \cite{Dubey2016} \cite{SantiagoAragonFedericoKuhlmann2015}  \\
            \hline
       IMSI catcher & GSM & USRP-N210/WBX USRP1& \cite{Hadzialic2014}
       \cite{Borgaonkar}
       \cite{Dabrowski2014} \\
          \hline
      Location leaks; denial of service & LTE & USRP B210 & \cite{Shaik2015} \\
          \hline
      Eavesdropping & ADS-B & RTL 2832U & \cite{TerrazasGonzalez2018} \\
          \hline 
      Eavesdropping & DECT  & USRP N210 RTL2832U & \cite{Sanchez2014} \\
          \hline
      Eavesdropping & Near Field Communication (NFC) & USRP N210  & \cite{Rong2016} \\
          \hline
      Protocols implementation & LoRaWan, BLE, IEEE802.11, IEEE802.15.4 & USRP E310 & \cite{Gavrila2018} \\
          \hline   
       TEMPEST  & Computer display & PXI-e 5665 USRP N210/WBX  & \cite{Elibol2012}
       \cite{Zhou2017} \\
          \hline 
      Side channel & Decryption AES-128 on 32 bits microcontroller & USRP2,RTL 2832U & \cite{Alakoca2017}
      \cite{Alakoca2016} \\
          \hline  
      Penetration Testing & Tactical Radio Networks & HackRF One  & \cite{Heinaaro2015} \\
          \hline
      Replay & RFID & USRP N210/SBX  & \cite{Han2015} \\
          \hline   
      Jamming & OFDM & NI USRP 2921 & \cite{Alakoca2017}
      \cite{Alakoca2016} \\
          \hline 
    
      vulnerabilities analysis on physical layer & LTE & USRP N210 & \cite{Rao2017} \\
          \hline
      \end{tabular}
      \\
      \caption{\label{tab:ejemplos_SDR}Cases of attacks, vulnerabilities, or cybersecurity analysis with SDR technology }
     
    \end{table*}

\subsubsection{\textbf{Spoofing attacks}}
This type of attack is also known as man in the middle, and seeks to impersonate some of the elements in communication by taking advantage of the weaknesses that may exist in the protocols. Impersonated technologies may be highly varied.\cite{Nguyen} shows the development of a drone detection system by analyzing the characteristics of wireless signals.\cite{Bresteau2018} performs an analysis on aeronautical communications technology ACARS (Aircraft Communication Addressing and Reporting System) and FANS1/A (Future Air Navigation System, using USRP B200 radios to generate messages, seeking to assess the impact of this type of attack on air safety.\cite{Kilcoyne2016} assesses the vulnerability of wireless communication used in automatic air pressure measurement systems on the wheels of vehicles,  \textit{Tire Pressure Monitoring System}, a technology required by the \textit{National Highway Trafic Safety Administration} in the United States. A USRP N210 was used to capture, analyze and impersonate the signals.\cite{Li2016} designed an attack against an indoor location system based on frequency modulations using a USRP B100 and GNU Radio platform, in order to evaluate the system's vulnerabilities. \cite{Schmidt2018} explores methods for mitigating GNSS(\textit{Global Navigation Satellite System}) signage spoofing attacks using SDR radios and algorithms implemented in FPGAs.

Experiments concerning GSM technology vulnerability analysis are presented in  \cite{Dubey2016} and \cite{SantiagoAragonFedericoKuhlmann2015} where USRP radios such as B210 or N210 are used to impersonate a GSM base radio. \cite{Hadzialic2014}, \cite{Dabrowski2014} and \cite{Borgaonkar} show the implementation of an IMSI-catcher or stingray device using a USRP B210 radio and OpenBTS \cite{OpenBTS}.\cite{Shaik2015} describes vulnerability tests for LTE network access protocols using USRP B200 and OpenLTE radio \cite{Wojtowicz}. The results show two different types of vulnerabilities evaluated: the first one makes it is possible to obtain the precise location of a device using GPS coordinates or True range multilateration from the signal intensity reported by the cell. The second test showed the generation of a (DoS) attack targeting an LTE device.

\subsubsection{\textbf{Eavesdropping attacks}}
\cite{Rong2016}  introduces a method to prevent eavesdropping attacks on Near Field Communication (NFC) devices. It employs a USRP N210 radio to generate variable signals as to amplitude, frequency, or phase and introduces additional bits to prevent an attacker from identifying NFC message sequences.\cite{Sanchez2014} explains an attack against a communication system with Digital Enhanced Cordless Telecommunication technology (DECT, using USRP SDR radios and RTL-SDR  \cite{RTL-SDR.COM} operating in a frequency range from 1880 to 1930 MHz. The developed system enabled protocol analysis and recovery of G-encoded voice signals G726.\cite{TerrazasGonzalez2018} describes the ADS-B- aerial positioning signals-demodulation process (\textit{Automatic Dependent Surveillance Broadcast} using an SDR-RTL device. \cite{BastilleNetworksa} shows the vulnerability of various types of wireless keyboards and mouses that work with uncoded wireless technologies, thus exposing passwords, personal information, bank details, etc.\cite{MatthewKnight2016} describes the process of decoding and analyzing the LoRa technology protocol using a USRP B210 radio, GNU Radio and Python.

\subsubsection{\textbf{Sidechannel attacks}}

\cite{Zhou2017} describes a TEMPEST-type side-channel attack for the reconstruction of images from radio frequency signals captured by a log periodic antenna and an SDR-USRP radio platform.\cite{Kuhn1998} describes the physical principles of a TEMPEST attack, where imaging on a monitor is analyzed. \cite{Elibol2012} presents the results of reconstructing the images of various types of monitors, reaching a distance of 46 meters, using a directive antenna and the National Instruments PXi-E 5665 hardware. \cite{NoriyukiMiura2016} shows the development of a sensor that detects a variations of electromagnetic fields near processing devices for the purpose of detecting possible attacks.

\subsubsection{\textbf{Jamming attacks}}

\cite{Alakoca2017} analyzes the effects of a jamming attack targeting an OFDM communication system, and analyzes metrics such as BER, MER and EVM. In the implementation of the testbed, NI USRP 2921 radios were used.\cite{Alakoca2016} an OFDMA system was implemented using USRP NI 2921 radios that allowed to measure performance against jamming attacks. \cite{Patwardhan2014} describes a method to mitigate the effect of a jamming attack using beamformig techniques; USRP2 and GNU Radio radios were used in the proof of concept.

\subsection{SDR as a cybersecurity tool }

The potential of Software-defined radio technology as a tool for analysis in cybersecurity is very high, especially when considering its cognitive possibilities. Cybersecurity analysis requires the use of techniques to assess the vulnerability of systems using procedures framed within ethical hacking. One of the techniques, known as systems penetration tests or pentest, allows to establish if there are vulnerabilities or if the defenses of a system are enough. The results of these tests help improve network security. One of the most commonly used tools for these procedures is Kali-Linux \cite{Lisitsa2018};  which is Linux distribution that includes support for some SDR platforms such as RTL, HackRF, UHD hardware and Funcube. The technologies included in the tool are limited to Wi-Fi, Bluetooth and NFC.
In \cite{Picod2014a} the authors describe the tool for penetration testing Scapy, developed in Python language, which allows capturing, decoding and building packages for protocols such as 802.15.4, ZWave and Wireless M-Bus using a USRP B210 device. \cite{Heinaaro2015} describes a set of tests to identify vulnerabilities in military radio networks. SDR USRP and HackRF One devices are used as radio platforms. Bastille Networks \cite{BastilleNetworks} is a company engaged in identifying risks in what they call the “Internet of radios”. For the development of their cybersecurity audit services, they have developed and patented their own tools to diagnose or control cybersecurity threats: Collaborative Bandit Sensing; Bayesian Device Fingerprinting and Distributed Tomographic Localization.These tools were developed from software-defined radio devices.
\cite{Visoottiviseth2017} describes the PENTOS tool, designed to perform penetration testing on IoT devices, but limited to Wi-Fi and Bluettoth technologies.In \cite{Sagduyu2017} the authors describe the design and implementation of the EZPro software tool, which enables the testing and evaluation of tactical communications against jamming, PU Emulation, protocol emulation and SSDF (\textit{Spectrum Sensing Data Falsification}) attacks.

\section{OVERVIEW AND DISCUSSION}
Figure \ref{fig:ecosistema} shows the magnitude of the IoT wireless ecosystem, operating frequencies, and standards. It shows several emerging technologies based on the 802.15.4 standard, such as: Wi-sun (802.15.4e), Thread, ZigBee NAN, ISA100.11a, OnRamp, Wireless Hart, MiWi, Positive Train ctrl (802.15.4p) and 802.15.4m. Another set of emerging  WiFi-based technologies was identified, to wit, 802.11ac, 802.11ad, 802.11ah (HaLow), 802.11p (WAVE), 802.11af (Super Wi-Fi or WhiteFi). One of the major fields of development and competition among manufacturers is LPWA (Low Powe Wide Area) networks, where there are two large unlicensed LPWA groups that group technologies such as: LoRa, SigFox, Telensa and OnRamp, and a second LPWA-licensed group encompassing mobile technologies based on LTE and GSM such as NB-IoT and EC-GSM-IoT.

Table \ref{tab:ADC-DAC} shows the technical characteristics of the most common software-defined radio platforms, taking into account the number of bits used in ADC and DAC converters, their frequency ranges, transmission and/or reception capabilities and sampling rates. SDR technology has evolved thanks to the development of converter manufacturing techniques, moving from sampling rates in the order of 200 [Kb/s] used in 1999 to rates of 200 [MS/s] in devices manufactured in 2016. 

SDR hardware is in full development; recent platforms such as CRISOM  \cite{Pervices}  have ADC sampling rates of 370.16 [MS/s] and 2500,16 [MS/s] for DAC, plus an ARM Cortex-A9 MP processor and FPGA-Arria V ST SOC. Other SDR platform developments integrate processing capabilities for working with deeplearning techniques, such as the AIR-T (Artificial Intelligence Radio - Transceiver) platform composed of a 2x2 SDR MIMO system with a 256-core Jetson TX2 GPU. Improvements in conversion rates and increased computing power make it possible to process the large flow of I/Q samples and the implementation of modern wireless communications standards.

Several types of wireless system attacks and vulnerabilities were identified. Figure \ref{fig:ataques_c1} presents a classification of the types of wireless attacks for SDR radios. Table \ref{tab:bastille} describes the most common wireless vulnerabilities, according to the company Bastille Networs.Table \ref{tab:ataques_cognitive}, describes the types of attacks against cognitive radio networks.Table \ref{tab:ataques_iot} presents IoT's various wireless vulnerabilities according to a three-tier model proposed by \cite{Frustaci2018}. The literature identifies the perception level as the most vulnerable, exposed to physical, impersonation, and routing and data traffic attacks, due to limitations in hardware, physical exposure of devices and heterogeneity of technologies. One of the efforts to develop solutions related to improving manufacturing processes is the DARPA-funded  \textit{Trusted Integrated Circuits} for the design of technologies to ensure the reliability of hardware used in military systems.

The variety of SDR platforms is increasingly wide, as seen in table \ref{tab:ADC-DAC}, which has allowed to democratize knowledge in software-defined radio technology. Table \ref{tab:ejemplos_SDR} shows various implementation cases of wireless attacks using software-defined radio technology.The most commonly used radio platforms for this purpose are the USRP. Among the most common implementations are: spoofing, Eavesdropping, side channel, jamming, replay. The implementation of pentesting tools that show the potential of SDR technology for the development of cybersecurity tools was identified as well.

The research challenges related to the security of IoT Technologies are manifold, considering the heterogeneous characteristics of wireless technologies and the lack of standards for manufacturing secure IoT devices.

Future systems based on SDR radios have great advantages due to their flexibility to adapt to new communication technologies, but this very flexibility poses great risks. It is still unclear how to prevent software in radios from being modified or what strategies are most convenient for the protection of software and hardware of devices.

SDR platforms can operate in a frequency range from DC to 6 GHz, a range where there is a high percentage of current communication technologies and also have reasonable processing capabilities. However, the potential of the technology lies in the software being developed and the processing techniques implemented, wherefore it is important to consider that the design of solutions aimed at cybersecurity in wireless systems is a complex issue that involves areas of knowledge such as digital signal processing, communications systems, networks and communication protocols, signal propagation, antennas and microwave systems, among others.

Cyber security challenges for the Internet of Things are so complex that software-defined radio technology itself cannot be considered as a solution. SDR technology can be seen as a tool that can be combined with cognitive techniques and intelligent techniques such as deeplearning, which can be easily adapted to rapid technological changes.In this sense, the most ambitious initiative undergoing development is the DARPA \textit{Spectrum Collaboration Challenge} \cite{DARPAa}, a challenge designed to be carried out over a period of three years (2017-2019).It seeks to apply artificial intelligence techniques to expand the capabilities of software-defined radio technology. It uses the \textit{Colosseum} infrastructure, which is a laboratory built at the Applied Physics Laboratory (APL) at Johns Hopkins University, composed of 256 USRP radios and a set of servers with GPU processing capabilities.

\section*{Acknowledgment}
Project INV-ING-2998 financed by UMNG

\bibliographystyle{IEEEtran}
\bibliography{main}

\end{document}